\documentstyle[prb,aps,multicol,epsf]{revtex}
\begin{document}
\draft
\title{Shallow Coulomb Gap and Weak Level-Level Correlations in a Deeply
Insulating Electron System with Interactions}
\author{Ya.~M.~Blanter$^{a}$ and M.~E.~Raikh$^{b}$}
\address{$^a$ D\'epartement de Physique Th\'eorique, Universit\'e de
Gen\`eve, CH-1211 Gen\`eve 4, Switzerland\\
$^b$ Department of Physics, University of Utah, Salt Lake City, Utah
84112, U.S.A.}
\date{\today}
\maketitle
\tighten
\begin{abstract}
We consider a system of two-dimensional electrons strongly localized
by disorder. Interactions create a gap in the {\em
average} tunneling density of states $\nu(E)$ at energies, $E$, close
to the Fermi level. We derive a system of self-consistent equations
for the correlators of {\em local} densities of occupied and empty
states. When either the interactions are screened by the
gate or the temperature, $T$, is high enough, so that the Coulomb gap
is shallow, the perturbative solution of the system yields
analytical expressions for the interaction-induced correction
$\delta\nu(E)$ and the level-level correlator.  We show that even with
short-range interactions, $\delta\nu (E)$ exhibits a singular energy
dependence at $T=0$. We also demonstrate that at high $T$ this energy
dependence is a {\em universal} function of the the ratio
$E/T$. Regarding the level correlations, we trace how the correlator
falls off as a function of spatial and energy separation between the
levels.  We also trace how the correlations vanish with increasing
$T$.  Our most noticeable observation is that for two close energies
the correlator changes sign from positive (attraction) at small
distances to negative (repulsion) at large distances.
\end{abstract}
\pacs{PACS numbers: 73.20.Dx, 71.23.An, 71.55.Jv}

\begin{multicols}{2}
\section{Introduction}

A distinctive difference between metallic and insulating systems lies
in statistics of energy levels. It is now a common knowledge that in
metallic diffusive samples, where electrons explore the entire volume
in course of their motion, energy levels obey Wigner-Dyson
statistics (for review, see Refs. \onlinecite{Efetov,Mirlin}). In the
deeply insulating regime electron states are localized, and can be
viewed as classical points randomly positioned in space and energy.
Thus, the levels obey the Poisson statistics. There is
presently no analytical theory describing how the Wigner-Dyson
statistics evolve into the Poisson one as the degree of disorder
grows. The most interesting critical regime, {\em i.e.} the level
statistics directly at the Anderson transition, was first addressed in
Refs. \onlinecite{Shklovskii1,Shklovskii2}, which generated an ongoing
stream of subsequent studies. In terms of experimentally observable
quantities, the level statistics correspond to the correlation in
tunnel conductances at different positions of the Fermi level.

It is also well known that electron-electron interactions affect
considerably the low-energy behavior of disordered systems. In the
diffusive regime, they generate a dip in the average tunneling density
of states at the Fermi surface. This dip manifests itself as the
zero-bias anomaly (for review, see Ref. \onlinecite{AA}). Interaction
effects are even more pronounced in the insulating regime
\cite{Pollak}. For Coulomb interactions between the electrons, the
tunneling density of states turns to zero as a power law in the
vicinity of the Fermi level, as argued by Efros and Shklovskii
\cite{Efros75}. This phenomenon, dubbed the Coulomb gap, also became a
subject of a big number of papers, which are too numerous to be cited
here. The works before 1985 are reviewed in
Refs. \onlinecite{Efros85,Pol-Ort85}; recent theoretical activity on
the classical Coulomb glass (aside from a long-standing problem of
hopping transport) was focused on the shape of the density of states
at zero\cite{Schreiber,Basylko} and finite temperatures with a finite
bandwidth of a bare energy distribution \cite{Pikus,Schrei-Pikus}.
Recent studies also addressed the dynamics of formation of the Coulomb
glass\cite{Pollak1,T-dep}.

Naively, one could think that the effect of interactions on the level
statistics in the classical case is only to suppress the density of
states, which enters as a factor in the Poisson distribution. Thus,
the resulting distribution might be expected to remain Poissonian. In
reality, however, even in a purely classical system interactions cause
deviations from the Poisson statistics of energy levels.
To be specific, let us address the following situation. Consider a
narrow energy strip close to the Fermi level. Without interactions,
the positions of sites with energies within the strip are completely
random. The prime effect of interactions is an overall depletion of the
strip (manifestation of the Coulomb gap).  Less obvious is the
question: to what extent, after the energies are modified by
interactions, the sites within the same strip ``feel'' each other?  In
other words, how are the {\em fluctuations} in their density
correlated? How does this correlation depend on the energy position of
the strip? How does it vanish with increasing temperature, {\em etc}.

Getting answers to the above questions by means of numerical
simulations is a much harder task than studying the shape
of the Coulomb gap. In view of this, an analytical approach to the
problem of level statistics in insulators with interactions is vitally
important.

In this paper, we develop an analytical theory of level statistics in
classical insulators with interactions.

Such a theory must apparently be based on one of the existing
analytical treatments of the Coulomb gap problem, which we review
below. The first approach is due to Efros \cite{selfconsis}. The basic
underlying assumption of this approach is that at zero temperature the
ground state of the system can be achieved by transpositions of
electrons within singly occupied pairs of localized states
(sites). Namely, for an isolated pair of sites with the energies
$\varepsilon_1$ and $\varepsilon_2$, separated by the distance
$\rho_{12}$, the work needed to transfer an electron from the occupied
site $1$ to the empty site $2$ is
\begin{equation} \label{work1}
\Delta_{12} = \varepsilon_2 - \varepsilon_1 - V(\rho_{12}),
\end{equation}
where $V(\rho)$ is the pair interaction potential. For the ground
state, this work must be positive for each pair of sites
$(1,2)$. Efros \cite{selfconsis} expresses the density of states
through the probability that all the conditions $\Delta_{12} > 0$ are
fulfilled. The advantage of the approach \cite{selfconsis} is that it
yields a closed equation for the density of states. The drawback is
that it neglects the possibility of reducing the total energy by
transpositions of electrons within the groups of three and more sites.
An alternative derivation of the self-consistent equation
\cite{selfconsis} is given in Ref. \onlinecite{selfconsis1}.

The formalism developed in Ref. \onlinecite{mogil} is also restricted
to transpositions of pairs. The difference between
Refs. \onlinecite{selfconsis} and \onlinecite{mogil} lies in the way
of pair counting. In Ref. \onlinecite{mogil} the pairs are ranged
according to the distance between the constituting sites $\rho_{12}$
(referred below as the {\em pair shoulder}). The theory
describes the evolution of the density of states with increasing
$\rho_{12}$. Consequently, the final result emerges as a solution
of a ``rate equation'' in the limit when all the transpositions
are carried out ($\rho_{12} \to \infty$). In contrast to
Ref. \onlinecite{selfconsis}, the formalism of Ref. \onlinecite{mogil}
preserves automatically the total number of occupied and empty
sites. By this virtue, the latter allows the generalization to finite
temperatures in a natural way: if the condition $\Delta_{12} > 0$ is
met for a certain pair, the transfer of the electron is not forbidden,
but happens with a probability $\exp(-\Delta_{12}/T)$.
Although the formalisms
of Refs. \onlinecite{selfconsis} and \onlinecite{mogil} are different,
the results they yield for the zero-temperature density of states
coincide.

Another approach was put forward by Johnson and Khmelnitskii
\cite{Johnson}. They start from the {\em exact} expression for the
partition function, expand it into the series of Feynman diagrams,
which they subsequently average over disorder term by term. In the
one-dimensional case they sum up the leading-order logarithmically
divergent diagrams. The result of the summation reproduces the
expression \cite{Raikh87} obtained in the framework of the approach of
Ref. \onlinecite{selfconsis}. For two- and three-dimensional cases ($D
= 2,3$), the results \cite{Efros75} for the density of states $\nu(E)
\propto \vert E \vert^{D-1}$ in the case of the Coulomb interactions
($V(\rho) \propto \rho^{-1}$) were reproduced in
Ref. \onlinecite{Johnson} by means of an $\epsilon$-expansion with
$\epsilon = D-1$.

Both approaches were designed to describe the {\em average} density of
states.

Below we study the effect of interactions on the {\em fluctuations} of
the density of states in disordered insulators, based on the
technique developed previously in Ref. \onlinecite{mogil}. In Section
\ref{geneq} we derive the general equations relating the
average density of states and the level-level correlation function for
an arbitrary pair interaction potential $V(\rho)$ {\em and finite
temperature}. Sections \ref{denssect} and \ref{statsect} are devoted to
the solution of these equations for the case when this potential can be
treated perturbatively. For this case we derive a general formula for
the density of states (Section \ref{denssect}) with an arbitrary
potential and apply it to the two particular geometries relevant for
experiment:  the 2D disordered insulator located above a planar gate,
and sandwiched between the two planar gates.  The perturbative
expression for level-level correlation function is derived and analyzed
in Section \ref{statsect}. Concluding remarks are presented in Section
\ref{concl}.

\section{General equations} \label{geneq}

The main subject of the study of level statistics is the level-level
correlation function, defined as follows,
\begin{eqnarray} \label{def1}
R^+ (\bbox{r}, E_1,E_2) \equiv \langle \nu^+(E_1, \bbox{R}) \nu^+(E_2,
\bbox{R} + \bbox{r}) \rangle_c \ ,
\end{eqnarray}
where the subindex $c$ stands for the irreducible part, $\langle AB
\rangle_c \equiv \langle AB \rangle - \langle A \rangle\langle B
\rangle$, and $\nu^+(E, \bbox{R})$ is the {\em local} density of {\em
occupied} states at the space point $\mbox{\bf R}$ and energy $E$,
\begin{equation} \label{def2}
\nu^+(E, \bbox{R}) = \sum_i n_i \delta( \bbox{R} - \bbox{R}_i)
\delta\left(\varepsilon_i - E \right).
\end{equation}
Here we consider a system of electrons which are localized at the
impurities randomly positioned at space points $\bbox{R}_i$, and
having (single-particle) energies $\varepsilon_i$; $n_i$ are the
occupation numbers, assuming the values $0$ or $1$. It is crucial to
emphasize that the single-particle energies $\varepsilon_i$ include
the interaction-induced shifts from all the surrounding sites
\cite{Efros85}. Note that the definition (\ref{def1}) differs from the
standard definition, commonly used in the diffusive regime (see {\em
e.g.} Refs. \onlinecite{Efetov,Mirlin}) in two respects. First, only
occupied states are taken into account. The existence of the Fermi
level (for which we choose $E = 0$) breaks the homogeneity in the
energy space, and thus Eq. (\ref{def1}) depends on the two energies
$E_1$ and $E_2$. Furthermore, the definition (\ref{def2}) implicitly
assumes that the particles are classical, {\em i.e.} the spatial
extent of the electron wave function is much smaller than the average
distance between the impurities. The level-level correlation function
(\ref{def1}) depends only on the difference $\bbox{r}$ of the space
arguments due to the global homogeneity. Because of this, the
averaging in Eq. (\ref{def1}) may be actually viewed as the
integration over $\bbox{R}$.

To construct the theory, it is convenient to introduce, in addition to
Eqs. (\ref{def1}), (\ref{def2}), the density of {\em empty} states,
\begin{equation} \label{def3}
\nu^-(E, \bbox{R}) = \sum_i (1 - n_i) \delta( \bbox{R} - \bbox{R}_i)
\delta\left(\varepsilon_i - E \right),
\end{equation}
and three more correlation functions,
\begin{eqnarray} \label{def4}
R^- (\bbox{r}, E_1,E_2) \equiv \langle \nu^-(E_1, \bbox{R}) \nu^-(E_2,
\bbox{R} + \bbox{r}) \rangle_c\ ,
\end{eqnarray}
\begin{eqnarray} \label{def5}
S (\bbox{r}, E_1,E_2) \equiv \langle \nu^+(E_1, \bbox{R}) \nu^-(E_2,
\bbox{R} + \bbox{r}) \rangle_c\ ,
\end{eqnarray}
and
\begin{eqnarray} \label{def6}
\tilde S (\bbox{r}, E_1,E_2) \equiv \langle \nu^-(E_1, \bbox{R})
\nu^+(E_2, \bbox{R} + \bbox{r}) \rangle_c\ ,
\end{eqnarray}
They are related by obvious symmetries, $R^{-} (\bbox{r},-E_1,-E_2) =
R^{+} (\bbox{r}, E_1,E_2)$, and $\tilde S (\bbox{r}, E_1, E_2) =
S (\bbox{r}, E_2,E_1)$. Thus, the knowledge of the functions $R^+$ and
$S$ is sufficient for the description of the level statistics.

Without interactions, the average density of states is constant
(we denote it with $\nu_0$), while
the fluctuations of the local density of states are Poissonian,
\begin{eqnarray} \label{indcond}
\langle \nu^{\pm} (E, \mbox{\bf R}) \rangle & = & \nu_0 f_F (\pm E); \\
R^{\pm} (\bbox{r}, E_1, E_2) & = & \nu_0 \delta(\bbox{r}) \delta(E_1 -
E_2) f_F (\pm E_1) f_F (\pm E_2); \nonumber \\
S(\bbox{r}, E_1, E_2) & = & \nu_0 \delta(\bbox{r}) \delta(E_1 - E_2)
f_F (E_1) [1 - f_F (E_2)]. \nonumber
\end{eqnarray}
Here $f_F(E) = [\exp(E/T) + 1]^{-1}$ is the Fermi function.
\begin{figure}
\narrowtext
{\epsfxsize=8.5cm
\centerline{\epsfbox{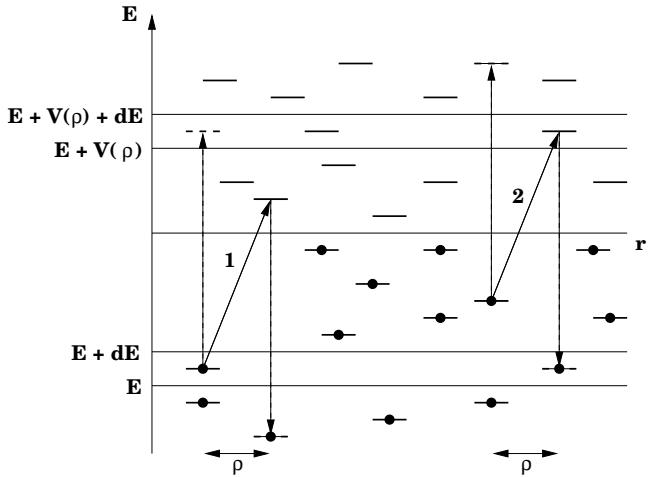}}}
\protect\vspace{0.5cm}
\caption{Transpositions of electrons between single and
occupied localized states, illustrating the derivation of
Eqs. (\ref{kin1}) and (\ref{kin2}). The states, characterized by
location in energy and space, are shown as dashes; occupied states are
marked by bullets. Processes 1 and 2 are explained in the text.} 
\label{fig1}
\end{figure}
\noindent

In the presence of interactions, Eqs. (\ref{indcond}) do not actually
describe the equilibrium state of the system. To find the equilibrium
state, we employ the approach of Ref. \onlinecite{mogil}. Within this
approach, the search for the equilibrium state is performed by
transpositions of electrons within pairs of singly-occupied states
({\em i.e.} the pairs $(i,j)$ with $n_i = 1$ and $n_j = 0$). To this
end, we introduce an auxiliary variable $\rho$ (the length of
the pair shoulder), and the functions $g^{\pm} (E, \bbox{R}, \rho)$.
These functions have the meaning of the local densities of
occupied/empty states, which are established after the transpositions
for all the pairs with the shoulders less than $\rho$ are already
carried out. For $\rho = 0$ the non-interacting values serve as an
initial condition; the limiting case $\rho \to \infty$ corresponds to
the equilibrium state of the system. The densities of states
$\nu^{\pm}$ (\ref{def2}) and $g^{\pm}$ are related via
\begin{equation} \label{dens001}
\nu^{\pm} (E, \bbox{R}) = \lim_{\rho \to \infty} g^{\pm} (E,
\bbox{R}, \rho).
\end{equation}

Now we derive the equation for the functions $g^{\pm}$. To be
specific, we consider now the two-dimensional case. The derivation
is best illustrated by Fig.~1. Consider the energy strip $(E, E +
dE)$. Imagine that all the transpositions with the shoulders shorter
than $\rho$ are already performed. As we increase the shoulder from
$\rho$ to $\rho + d\rho$, the single-particle energies change due to
transpositions, and two processes take place: (i) a certain number of
occupied sites leave the energy strip (Process 1 in Fig.~1); (ii) as a
result of the transpositions, the empty sites with the single-particle
energies within the strip $(E + V(\rho), E + dE + V(\rho))$ get
occupied and arrive into the strip $(E, E + dE)$ (Process 2 in
Fig.~1). The net change of content of the energy strip is responsible
for the change $\delta g^+ (E, \bbox{R}, \rho)$ of the density of
states. Thus, the evolution of $g^+$ with $\rho$ is described by an
equation which is similar to the rate equation of the kinetic theory,
with the processes (i) and (ii) corresponding to out-scattering and
in-scattering terms in the collision integral, respectively. The
actual form of this equation is as follows,
\begin{eqnarray} \label{kin1}
& & \frac{\partial g^+ (E, \bbox{R}, \rho)}{\partial \rho} =
-\frac{\rho}{2} \int d\bbox{n} \Biggl\{ g^+ (E, \bbox{R}, \rho)
\Biggr.\\
& \times & \left[ \int_{-\infty}^{E + V(\rho)} dE' g^- (E', \bbox{R} +
\bbox{n}\rho, \rho) + \int_{E + V(\rho)}^{\infty} dE' \right.
\nonumber\\
& \times & \left. \left. g^- (E', \bbox{R} + \bbox{n}\rho, \rho) \exp
\left( -\frac{E' - E - V(\rho)}{T}
\right) \right] \right. \nonumber \\
& - & \left. g^- (E + V(\rho), \bbox{R}, \rho) \left[
\int_{E}^{\infty} dE' g^+ (E', \bbox{R} + \bbox{n}\rho, \rho)
\right. \right. \nonumber\\
& + & \left. \left. \int_{-\infty}^E dE' g^+ (E', \bbox{R} +
\bbox{n}\rho, \rho) \exp \left( -\frac{E - E')}{T} \right) \right]
\right\}, \nonumber
\end{eqnarray}
where $\bbox{n}$ is a unit vector, and $\int d\bbox{n} = 2\pi$.

Let us establish the correspondence between the r.h.s. of
Eq. (\ref{kin1}) and the ``scattering'' processes shown in Fig.~1. The
first two terms (first square bracket) in the r.h.s. describe the
scattering out, {\em i.e.} departure of levels from the energy strip
$(E, E+dE)$ due to the transpositions (Process 1). Indeed, this
bracket is multiplied by the factor $g^+$, describing the electrons
which have been transferred. The bracket itself contains the function
$g^-$ as an integrand, which accounts for empty levels {\em to} which
the electrons are transferred. The argument $\bbox{R} + \bbox{n}\rho$ of
this function takes into account that the transpositions occur
within the pairs with the shoulder $\rho$. Furthermore, in the
integration over the energies of all available sites, it is taken into
account that the transpositions with negative work $\Delta (E,E') = E'
- E - V(\rho)$ happen with the probability one, while for $\Delta
(E,E') > 0$ the probability is given by the Boltzmann factor. The last
two terms in the same way describe the scattering in (Process 2). It
is important that Eq. (\ref{kin1}) {\em is not averaged} over the
position $\bbox{R}$, and thus carries the information about the
local fluctuations of the density of states.

Similarly, the ``kinetic'' equation for $g^-$ takes the form
\begin{eqnarray} \label{kin2}
& & \frac{\partial g^- (E, \bbox{R}, \rho)}{\partial \rho} =
-\frac{\rho}{2} \int d\bbox{n} \Biggl\{ g^- (E, \bbox{R}, \rho) \Biggr.
\\
& \times & \left[ \int_{E - V(\rho)}^{\infty} dE' g^+ (E', \bbox{R} +
\bbox{n}\rho, \rho) + \int_{-\infty}^{E - V(\rho)} dE' \right.
\nonumber\\
& \times & \left. g^+ (E',
\bbox{R} + \bbox{n}\rho, \rho) \exp \left( -\frac{E - E' - V(\rho)}{T}
\right) \right] \nonumber \\
& - & \left. g^+ (E - V(\rho), \bbox{R}, \rho) \left[
\int_{-\infty}^E dE' g^- (E', \bbox{R} + \bbox{n}\rho, \rho)
\right. \right. \nonumber\\
& + & \left. \left. \int^{\infty}_E dE' g^- (E', \bbox{R} +
\bbox{n}\rho, \rho) \exp \left( -\frac{E' - E)}{T} \right) \right]
\right\}. \nonumber
\end{eqnarray}

Now we use the equations (\ref{kin1}) and (\ref{kin2}) to obtain the
coupled equations for the average density of states and the
level-level correlation functions. The analogy between
Eqs. (\ref{kin1}) and (\ref{kin2}) and the kinetic theory of gases
suggests the following course of action. We split the functions
$g^{\pm}$ into average (over the space) and fluctuating parts,
\begin{equation} \label{split1}
g^{\pm} (E, \bbox{R}, \rho) = \nu_{\rho}^{\pm} (E) + \delta
g^{\pm} (E, \bbox{R}, \rho),
\end{equation}
with $\langle \delta g^{\pm} \rangle = 0$. First, we average
Eq. (\ref{kin1}). After this procedure, the rate of change of the
average value of $g^+$ is expressed through $\nu_{\rho}^+$,
$\nu_{\rho}^-$, and the correlator of the fluctuations $\delta
g^{\pm}$,
\begin{eqnarray} \label{basic1}
\frac{\partial \nu_{\rho}^+ (E)}{\partial \rho} & = & -\pi \rho
\left\{ \nu_{\rho}^+ (E) \int dE' \psi[E + V(\rho) - E'] \nu_{\rho}^-
(E') \right. \nonumber \\
& - & \left. \nu_{\rho}^- [E + V(\rho)] \int dE' \psi(E' - E)
\nu_{\rho}^+ (E') \right. \\
& + & \int dE' \psi[E + V(\rho) - E'] S_{\rho} (\rho, E, E')
\nonumber \\
& - & \left. \int dE' \psi(E' - E) S_{\rho} [\rho, E', E + V(\rho)]
\right\}, \nonumber
\end{eqnarray}
where we have introduced the function
\begin{eqnarray} \label{psifun}
\psi(E) = \left\{ \matrix{ 1, & E > 0 \cr
e^{E/T}, & E < 0} \right. \ \ \ .
\end{eqnarray}
The quantity $S_{\rho}$, which enters the r.h.s.,
is the cross-correlator of the fluctuations $\delta g^+$ and $\delta
g^-$,
\begin{equation} \label{defcor1}
S_{\rho} (r, E_1, E_2) \equiv \langle \delta g^+ (E_1, \bbox{R}, \rho)
\delta g^- (E_2, \bbox{R} + \bbox{r}, \rho) \rangle,
\end{equation}
which, due to the spatial homogeneity, depends only on the distance
$r$ between the two points.

Averaging of Eq. (\ref{kin2}) yields an analogous equation relating
$\nu_{\rho}^-$ to $\nu_{\rho}^+$ and the correlator
$S_{\rho}$. However, we observe that the electron-hole symmetry, which
applies for all {\em average} quantities, implies $\nu^-_{\rho} (E) =
\nu^+_{\rho} (-E)$. Thus, this second equation is unnecessary.

If we neglect the fluctuation-induced contribution to the density of
states (terms with $S_{\rho}$ in the r.h.s.), Eq. (\ref{kin2}) turns
into a closed equation for $\nu^+_{\rho}$ which coincides with that
obtained previously in Ref. \onlinecite{mogil}.

The equation (\ref{basic1}) needs to be supplemented by two more
equations describing the evolution of the correlation functions
$S_{\rho}$ and $R^{\pm}_{\rho}$, defined as
\begin{equation} \label{defcor2}
R^{\pm}_{\rho} (r, E_1, E_2) \equiv \langle \delta g^{\pm} (E_1,
\bbox{R}, \rho) \delta g^{\pm} (E_2, \bbox{R} + \bbox{r}, \rho)
\rangle.
\end{equation}
To derive these equations, we first subtract from Eq. (\ref{kin1}) for
$\partial g^+ (E_1, \bbox{R}, \rho)/\partial \rho$ its average,
Eq. (\ref{basic1}), and call the resulting equation $A$.
In the same way, we subtract from Eq. (\ref{kin2}) for $\partial g^+
(E_2, \bbox{R} + \bbox{r}, \rho)/\partial \rho$ its average and call
the resulting equation $B$. Following the general approach of the
kinetic theory of gases, we now multiply $A$ by $\delta g^- (E_2,
\bbox{R} + \bbox{r}, \rho)$, multiply $B$ by $\delta g^+ (E_1,
\bbox{R}, \rho)$, add them up and take the average, disregarding
triple correlations (see below Section \ref{concl}). As a result, we
obtain 
\end{multicols}
\widetext
\vspace*{-0.2truein} \noindent \hrulefill \hspace*{3.6truein}
\begin{eqnarray} \label{basic2}
& & \frac{\partial S_{\rho} (r, E_1, E_2)}{\partial \rho} \nonumber \\ 
& = & -\frac{\rho}{2} \left\{ 2\pi S_{\rho} (r, E_1, E_2) \int dE'
\psi[E_1 - E' + V(\rho)] \nu^-_{\rho} (E') \right. -  \left. 2\pi
R_{\rho}^- [r, E_2, E_1 + V(\rho)] \int dE' \psi[E' - E_1]
\nu^+_{\rho} (E') \right. \nonumber \\ 
& + & \left. \nu^+_{\rho} (E_1) \int d\bbox{n} \int dE' \psi [E_1 - E'
+ V(\rho)] R_{\rho}^- \left[\vert \bbox{r} + \bbox{n}\rho \right\vert,
E', E_2] \right. \nonumber \\
& - & \left. \nu^-_{\rho} [E_1 + V(\rho)] \int d\bbox{n}
\int dE' \psi (E'- E_1) S_{\rho} \left[\vert \bbox{r} + \bbox{n}\rho
\vert, E', E_2 \right] \right. \nonumber \\
& + & \left. 2\pi S_{\rho} (r, E_1, E_2) \int dE' \psi[E' -
E_2 + V(\rho)] \nu^+_{\rho} (E') \right. - \left. 2\pi R_{\rho}^+ [r,
E_2 - V(\rho), E_1] \int dE' \psi(E_2 - E') \nu^-_{\rho} (E')
\right. \nonumber \\ 
& + & \left. \nu^-_{\rho} (E_2) \int d\bbox{n} \int dE' \psi [E' - E_2
+ V(\rho)] R_{\rho}^+ [\vert \bbox{r} + \bbox{n}\rho \vert, E', E_1]
\right. \nonumber \\
& - &  \left. \nu^+_{\rho} [E_2 - V(\rho)] \int d\bbox{n} \int dE' \psi
(E_2- E') S_{\rho} [\vert \bbox{r} + \bbox{n}\rho \vert, E_1, E']
\right\}.  
\end{eqnarray}
\hspace*{3.6truein}\noindent \hrulefill 
\begin{multicols}{2}
\noindent

For the second equation, we ultimately multiply $A$ by $\delta g^+
(E_2, \bbox{R} + \bbox{r}, \rho)$, multiply $B$ by $\delta g^- (E_1,
\bbox{R}, \rho)$, add up and take the average. The resulting equation
reads
\end{multicols}
\widetext
\vspace*{-0.2truein} \noindent \hrulefill \hspace*{3.6truein}
\begin{eqnarray} \label{basic3}
& & \frac{\partial R^+_{\rho} (r, E_1, E_2)}{\partial \rho} \nonumber
\\ & = &
-\frac{\rho}{2} \left\{ 2\pi R^+_{\rho} (r, E_1, E_2) \int dE' \psi[E_1 -
E' + V(\rho)] \nu^-_{\rho} (E') - 2\pi S_{\rho} [r, E_2, E_1 +
V(\rho)] \int dE' \psi[E' - E_1] \nu^+_{\rho} (E') \right. \nonumber \\
& + & \left. \nu^+_{\rho}
(E_1) \int d\bbox{n} \int dE' \psi [E_1 - E' + V(\rho)] S_{\rho}
[\vert \bbox{r} + \bbox{n}\rho \vert, E_2, E'] \right. \nonumber \\
& - &  \left. \nu^-_{\rho} [E_1 + V(\rho)] \int d\bbox{n} \int dE' \psi
(E'- E_1) R^+_{\rho} [\vert \bbox{r} + \bbox{n}\rho \vert, E_2, E']
\right. \nonumber \\
& + & \left. 2\pi R^+_{\rho} (r, E_1, E_2) \int dE' \psi[E_2 -
E' + V(\rho)] \nu^-_{\rho} (E') - 2\pi S_{\rho} [r, E_1, E_2 +
V(\rho)] \int dE' \psi(E' - E_2) \nu^+_{\rho} (E') \right. \nonumber
\\ 
& + & \left. \nu^+_{\rho} (E_2) \int d\bbox{n} \int dE' \psi [E_2 - E'
+ V(\rho)] S_{\rho} [\vert \bbox{r} + \bbox{n}\rho \vert, E_1, E']
\right. \nonumber \\
& - &  \left. \nu^-_{\rho} [E_2 + V(\rho)] \int d\bbox{n} \int dE' \psi
(E'- E_2) R^+_{\rho} [\vert \bbox{r} + \bbox{n}\rho \vert, E_1, E']
\right\}. 
\end{eqnarray}
\hspace*{3.6truein}\noindent \hrulefill 
\begin{multicols}{2}
\noindent

Equations (\ref{basic1}), (\ref{basic2}), and (\ref{basic3}) describe
the evolution of the functions $\nu_{\rho}$, $S_{\rho}$, and
$R^+_{\rho}$ with increasing $\rho$. Due to the
electron-hole symmetry, $R_{\rho}^-(r, E_1, E_2) = R_{\rho}^+(r, -E_1,
-E_2)$, and thus the system of the equations (\ref{basic1}),
(\ref{basic2}), and (\ref{basic3}) is {\em closed}. For $\rho = 0$,
when no transpositions have yet been performed, the system corresponds
to the equilibrium state of non-interacting localized electrons. Thus,
Eqs. (\ref{indcond}) serve as initial conditions at $\rho = 0$,
\begin{eqnarray} \label{indcond1}
\nu^+_{\rho} (E) \vert_{\rho = 0} & = & \nu_0 f_F (E);
\nonumber \\
R^+_{\rho} (r, E_1, E_2) \vert_{\rho = 0} & = &
 \frac{\nu_0}{2\pi r}  \delta(r) \delta(E_1 - E_2) f_F (E_1) f_F
 (E_2); \nonumber \\
S_{\rho} (r, E_1, E_2) \vert_{\rho = 0} & = & \frac{\nu_0}{2\pi r}
 \delta(r) \delta(E_1 - E_2) \nonumber \\
& \times & f_F (E_1) [1 - f_F (E_2)].
\end{eqnarray}

To obtain equilibrium properties of interacting system within this
approach, one must solve Eqs. (\ref{basic1}), (\ref{basic2}), and
(\ref{basic3}). The average density of the occupied states $\nu^+ (E)$
and the correlation functions $R^+ (r, E_1, E_2)$ and $S (r, E_1, E_2)$
are then obtained as the limit $\rho \to \infty$ of the functions
$\nu_{\rho}^+ (E)$, $R^+_{\rho} (r, E_1, E_2)$, and $S_{\rho} (r, E_1,
E_2)$, respectively.

\section{Density of states} \label{denssect}

The cumbersome equations (\ref{basic1}), (\ref{basic2}), and
(\ref{basic3}) contain information about the statistical properties of
the equilibrium state of the system for an arbitrary interaction
potential $V$. In this and the next Sections our goal is to study {\em
weak} deviations of the level statistics from the Poisson
distribution, and of the average density of states from its
non-interacting value $\nu_0$. We solve this problem perturbatively
and demonstrate that the desired deviations appear already in the
first successive approximation.

To this end, we notice that the right-hand sides in Eqs.
(\ref{basic1}), (\ref{basic2}), and (\ref{basic3}) are effectively
small due to the energy integration. Thus, in the leading order
approximation it is sufficient to put the initial conditions
(\ref{indcond1}) into the right-hand sides of these equations, and
find the deviations from the initial conditions by integrating them
over $\rho$.

For the density of states, inserting Eq. (\ref{indcond1}) into the
right-hand side of Eq. (\ref{basic1}) and performing the energy
integration, we obtain the following equation for $\nu^+_{\rho}$,
\begin{eqnarray} \label{hightemp1}
\frac{\partial \nu^+_{\rho} (E)}{\partial \rho} & = & -\pi\rho\nu_0^2
T \nonumber \\
& \times & \left\{ \frac{\exp \frac{E + V(\rho)}{T}}{1 + \exp
\frac{E}{T}} \ln \left( 1 + \exp \frac{-E-V(\rho)}{T} \right) \right.
\nonumber \\ & + & \left. \frac{1}{1 + \exp \frac{E}{T}} \ln \left( 1 +
\exp \frac{E+V(\rho)}{T} \right) \right. \nonumber \\
& - & \left. \frac{\exp \frac{-E}{T}}{1 + \exp
\frac{-E-V(\rho)}{T}} \ln \left( 1 + \exp \frac{E}{T} \right)
\right. \nonumber \\
& - & \left. \frac{1}{1 + \exp \frac{-E-V(\rho)}{T}} \ln \left( 1 +
\exp \frac{-E}{T} \right) \right\},
\end{eqnarray}
valid for an arbitrary temperature\cite{foot1}. The total average
density of states is found as $\nu(E) = \nu^+ (E) + \nu^-(E)$, where
$\nu^+(E)$ is the solution of Eq. (\ref{hightemp1}) with the boundary
condition (\ref{indcond1}), taken for $\rho \to \infty$. Due to the
particle-hole symmetry, this becomes $\nu(E) = \nu^+ (E) + \nu^+
(-E)$.

Although Eq. (\ref{hightemp1}) can not be integrated for an arbitrary
temperature, in two important limiting cases results may be
obtained in a closed form.

\subsection{$T = 0$}

\subsubsection{General results}

For zero temperature, Eq. (\ref{hightemp1}) simplifies to
\begin{equation} \label{denszero0}
\frac{\partial \nu^+_{\rho} (E)}{\partial \rho} = -\pi\rho\nu_0^2
\left[ 2E + V(\rho) \right] \theta(-E) \theta \left[ E + V(\rho)
\right].
\end{equation}
Integrating Eq. (\ref{denszero0}), we find for the true average
density of states
\begin{equation} \label{dens1}
\nu (E) = \nu_0 \left[ 1 + \pi\nu_0 \vert E \vert \rho_E^2 - \pi
\nu_0 \int_0^{\rho_E} d\rho \rho V(\rho) \right],
\end{equation}
where $\rho_E$ is the solution of the equation $V(\rho) = \vert E
\vert$.

It is remarkable that although the right-hand side of the ``kinetic
equation'' Eq. (\ref{denszero0}) vanishes after being integrated over
$dE$, thus providing the independence of the total number of levels
on the interactions, the solution (\ref{dens1}) does not obey this
relation. To realize this, we recall that for $\rho \to 0$ each
realistic potential has a Coulomb form, $V(\rho) = e^2/\kappa
\rho$. Then it is straightforward to check the relation $\int dE
[\nu(E) - \nu_0] = -\pi \nu_0^2e^4/2\kappa^2$, in which the right-hand
side assumes a finite value due to the short-distance behavior of
$V(\rho)$. This seeming inconsistency has the following
resolution. Let us introduce a cut-off distance $\rho_{min}$ (minimal
separation between the sites) which is much smaller than the
characteristic decay length, $d$, of the potential $V(\rho)$.
Then for energies $E > e^2 (\kappa \rho_{min})^{-1}$ we have
$\nu (E)$= $\nu_0$. With finite $\rho_{min}$
the conservation of the number of states, $\int dE [\nu(E)
- \nu_0] = 0$, gets restored. This restoration occurs due the
fact that within a {\em wide} interval of high energies
[$e^2(\kappa \rho_{min})^{-1}\ll E \ll e^2 (\kappa d)^{-1}$]
the deviation  $\nu (E) - \nu_0$
exhibits a {\em shallow} maximum. While the height of this
maximum vanishes with  $\rho_{min} \to 0$, the area under
the maximum remains finite.

Conceptually more interesting is the low-energy behavior of the
density of states. For $E = 0$, we obtain $\nu(0) = \nu_0 [ 1 -
\pi\nu_0 \int_0^{\infty} d\rho \rho V(\rho)] < \nu_0$. Thus, the
average density of states at the Fermi level is suppressed by
interactions. This effect is a precursor of the Coulomb gap.

The perturbative treatment is only valid in the case when the
correction to the non-perturbed density of states $\nu_0$, given by
Eq. (\ref{dens1}), is small. Thus, the expansion parameter in the
theory is $\nu_0 d^2 V(d)$. If the potential $V(\rho)$
falls off at large distances slower than $\rho^{-2}$, the integral
$\int_0^{\infty} d\rho \rho V(\rho)$ diverges, manifesting that these
potentials (including the important case of the Coulomb potential) are
not amenable to the perturbative treatment at $T = 0$.

Now we apply the general result Eq. (\ref{dens1}) to two specific
realizations of short-range interaction potentials.

\subsubsection{2D system with a single gate}

In the presence of a planar gate at a distance $d$ from the 2D
electron gas, the Coulomb potential is modified due to the image
charge,
\begin{equation} \label{int21}
V(\rho) = \frac{e^2}{\kappa} \left( \frac{1}{\rho} -
\frac{1}{\sqrt{\rho^2 + 4d^2}} \right),
\end{equation}
where $\kappa$ is the static dielectric constant. In this case the
decay length is $d$; for larger distances $V(\rho)$ falls off as
$\rho^{-3}$. With interaction (\ref {int21}) the integration in
Eq. (\ref {dens1}) can be carried out in a closed from. However, the
radius $\rho_E$ is now a solution of the fourth power algebraic
equation. Then it is convenient to introduce the dimensionless energy
$\tilde E = 2d\kappa E/e^2$ and present the result for the density of
states as follows, 
\begin{equation}
\label{dens21} \nu(E)-\nu_0 = - \frac{2\pi\nu_0^2 e^2 d}{\kappa}
\frac{ (y(\vert \tilde E \vert) - 1 )^2}{y^2 (\vert \tilde E \vert) +
1},
\end{equation}
where
the function $y (\tilde E)$ is the positive solution of the equation
$y^4 + (4/\vert \tilde E \vert) y^3 - 1 = 0$. For $E = 0$ we obtain
$\nu(0) - \nu_0 = - 2\pi\nu_0^2 d e^2/\kappa$. In the vicinity of the
Fermi surface the interaction-induced correction has a singularity,
\begin{equation}
\label{singul21} \nu(E) - \nu(0) = 2\pi \nu_0^2 \left( \frac{2 e^2
d^2}{\kappa} \right)^{2/3} \vert E \vert^{1/3}.
\end{equation}
The energy dependence of the density of states calculated from
Eq. (\ref{dens21}) is shown in Fig.~2.

\begin{figure}
\narrowtext
{\epsfxsize=8.0cm
\centerline{\epsfbox{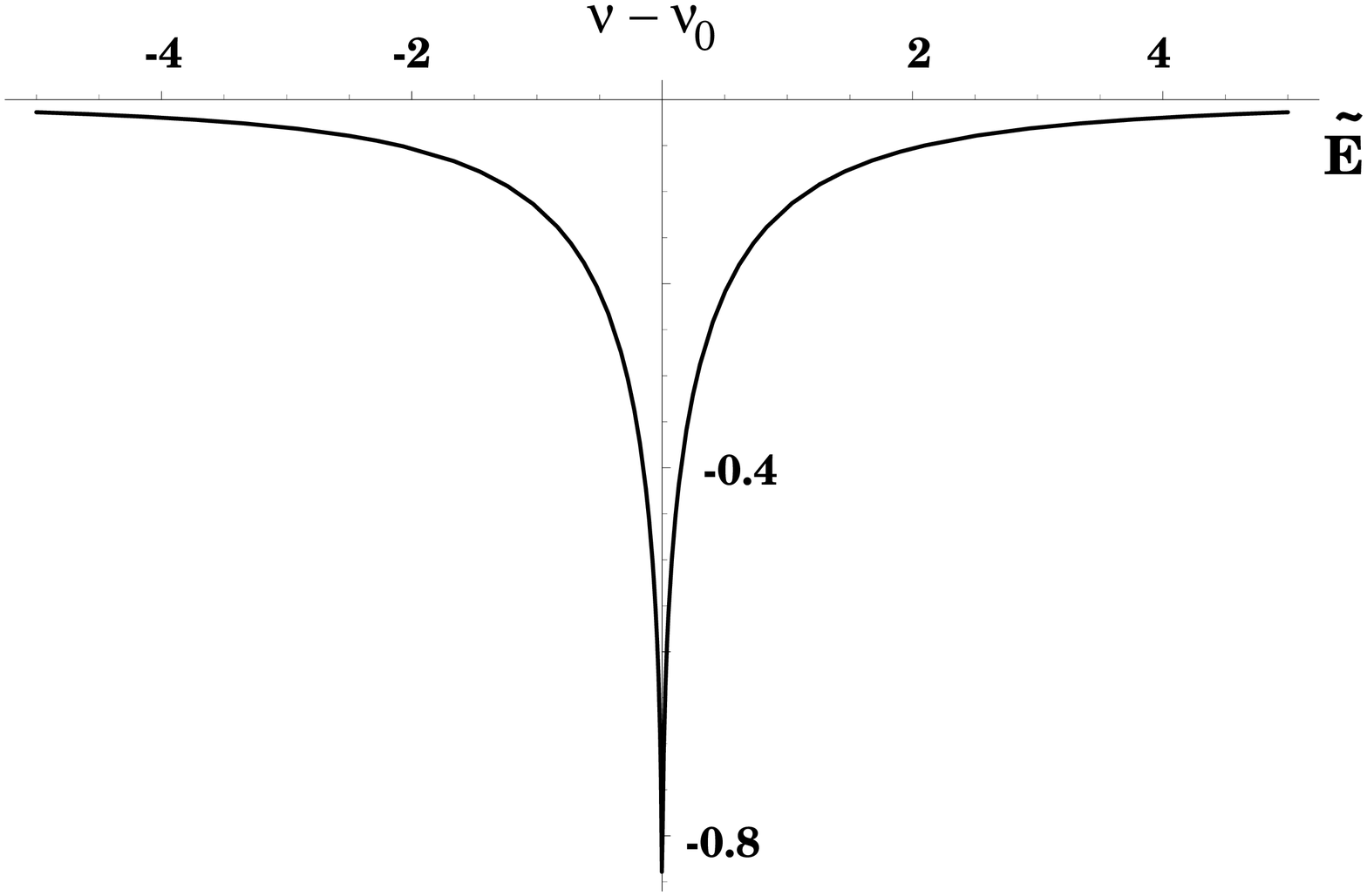}}}
\protect\vspace{0.5cm}

\caption{Density of states (\ref{dens21}) for a 2D system
with a single gate in units of $2\pi\nu_0^2e^2d/\kappa$ as a function
of the dimensionless energy $\tilde E$ at zero temperature.} 
\label{fig2}
\end{figure}
\noindent

\subsubsection{2D system between two gates}

We consider now a planar electron system located between two infinite
planar gates. Denote with $d_1$ and $d_2$ the distances to the
upper and lower gates. For large $\rho \gg (d_1+d_2)$ it  can be shown
that the interaction $V(\rho)$ takes the form
\begin{eqnarray} \label{gates}
V(\rho)& = & \frac{e^2}{\pi\kappa \sqrt{2\rho(d_1+d_2)}
}\cos^2\Biggl[\frac{\pi(d_1-d_2)} {2(d_1+d_2)}\Biggr] \nonumber \\
& \times & \exp\Bigl(-\frac{\pi\rho}{d_1+d_2}\Bigr)=
V_0\Bigl(\frac{a}{\rho}\Bigr)^{1/2} \exp(-\rho/a),
\end{eqnarray}
where $a=(d_1+d_2)/\pi$ is the interaction radius; the parameter $V_0$
is defined by the second equality in (\ref {gates}).
The asymptotic expression (\ref {gates}) is only
suitable to describe energies
$\vert E \vert \ll V_0$. In this case, keeping the two leading terms,
we write
\begin{displaymath}
\rho_E = a \left[ \ln \frac{V_0}{\vert E \vert} - \frac{1}{2} \ln
\Bigl(\ln
\frac{V_0}{\vert E \vert}\Bigr) \right].
\end{displaymath}
The leading order in $V_0/\vert E \vert$ terms in the density of
states are found from Eq. (\ref{dens1}),
\begin{eqnarray} \label{dens3}
& & \nu(E) - \nu_0 = \pi\nu_0^2 a^2 \left[ -\frac{\pi^{1/2}}{2} V_0
\right. \\
& + & \left. \vert E \vert \ln \frac{V_0}{\vert E \vert} \left( \ln
\frac{V_0}{\vert E \vert} - \ln \Bigl(\ln \frac{V_0}{\vert E \vert}
\Bigr) + O(1) \right) \right]. \nonumber
\end{eqnarray}

The singularity around the Fermi level is of the type
\begin{equation} \label{singul31}
\nu(E) - \nu(0) \propto \vert E \vert \ln^2 (V_0/\vert E \vert).
\end{equation}
It is remarkable that the singularity survives despite the fast decay
of the potential $V(\rho)$ with distance. Moreover, since the behavior
around $E = 0$ is determined exactly by this long-distance decay, the
singularity of the form (\ref{singul31}) is universal and
characteristic of any potential which falls off exponentially at large
distances. In any realistic experimental setup the sample is always
surrounded by a number of gates, and may be characterized by several
length scales. The long-{$\rho$} (beyond the longest scale) behavior of
this potential is always exponential. Thus, we conclude that the
$\vert E \vert \ln^2 \vert E \vert$ singularity at {\em small} $E$ is
generic.

\subsection{High temperatures}

We assume now that the temperature is higher
than the potential $V(\rho)$ for any $\rho$. In particular, this means
that the potential is finite everywhere. Expanding
Eq. (\ref{hightemp1}) in $V(\rho)/T$
up to the second order (first-order terms cancel out)
and subsequently integrating over
$\rho$, we find the high-temperature expression for the density of
states,
\begin{equation} \label{denshigh1}
\nu(E) =  \nu_0 + \frac{\nu_0^2{\cal L}}{T} F (E/T),
\end{equation}
where we have introduced the function
\begin{equation} \label{func1}
F (x) = \frac{\pi}{2 \cosh^2 (x/2)} \left\{ \frac{1}{2} - \frac{\ln
\left( 1 + e^x \right)}{1 + e^x} - \frac{\ln \left( 1 + e^{-x}
\right)}{1 + e^{-x}} \right\},
\end{equation}
and the parameter
\begin{equation}\label{L}
{\cal L}= \int_0^{\infty} d\rho \rho V^2 (\rho).
\end{equation}
It is straightforward to check that $\int dx F (x) = 0$, which
automatically ensures the conservation of the total number of states.
The effective upper limit in the integral (\ref {L}) is $\rho\sim d$,
where $d$ is a characteristic radius of the interaction potential.
However, the integral diverges logarithmically at small $\rho$, since
at short distances the interaction is always Coulomb,  $V(\rho) =
e^2/\kappa\rho$. In fact, this divergence is an artifact of the
expansion  of (\ref{hightemp1}) with respect to $V(\rho)/T$, which is
valid only for $\rho \gg \rho_T= e^2/\kappa T$.  Thus, $\rho_T$ serves
as an ultraviolet cut-off in the integral (\ref {L}). Taking this into
account, we obtain
\begin{equation}\label{L1}
{\cal L}=\frac{e^4}{\kappa^2} \ln \Biggl(\frac{\kappa d T}{e^2}
\Biggr).
\end{equation}
\begin{figure}
\narrowtext
{\epsfxsize=8.0cm
\centerline{\epsfbox{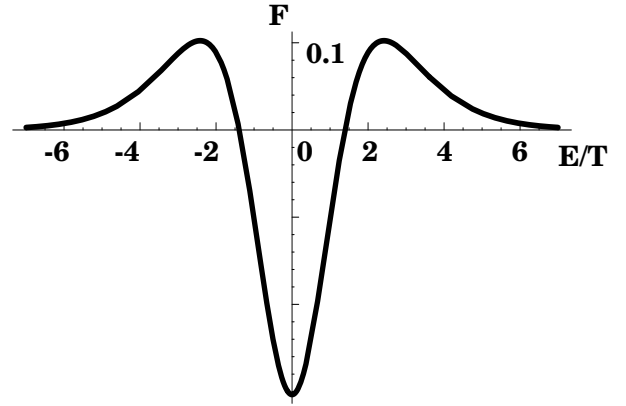}}}
\protect\vspace{0.5cm}

\caption{The universal function $F(E/T)$ (Eq. (\ref{func1})).} 
\label{fig3}
\end{figure}
\noindent
The actual form of the interaction potential affects only a
numerical factor under the logarithm. So does the contribution
from the short distances $\rho \lesssim \rho_T$. The condition
of validity of the high-temperature asymptotics (\ref {denshigh1})
is $T \gg V(d)\sim e^2/\kappa d$, {\em i.e.} the argument
of logarithm
in (\ref {L1}) should be much bigger than one. Another condition
is that the magnitude of the interaction-induced correction should
be smaller than $\nu_0$. The latter can be expressed as
$T\gg \nu_0e^4/\kappa^2$. This condition has a transparent
physical interpretation: $T$ should be bigger than the width of
the Coulomb gap at zero temperature.

It is  remarkable that the density of states (\ref{denshigh1})
depends on energy only in combination $E/T$, and thus the result is
expressed in terms of the {\em universal} function $F$, shown in
Fig.~3. The asymptotic expressions for $\nu(E)$ are
\begin{eqnarray} \label{ashigh1}
\nu(E) = \nu_0 - \frac{\pi\nu_0^2{\cal L}}{4T} \left( 2 \ln 2 - 1 - \frac{E^2
\ln 2}{2T^2} \right)
\end{eqnarray}
for $\vert E \vert \ll T$, and
\begin{eqnarray} \label{ashigh2}
\nu(E) = \nu_0 + \frac{\pi\nu_0^2{\cal L}}{T} e^{-\vert E
\vert/T}
\end{eqnarray}
for $\vert E \vert \gg T$.
As it is  seen from Fig.~3, the maxima of $\nu(E)$ correspond to
$E=\pm 2.2T$.

\section{Level statistics} \label{statsect}

\subsection{Singular part}

The level-level correlation function is defined by
\begin{eqnarray} \label{corpert1}
& & K(r, E_1, E_2) \equiv \langle \Delta\nu(E_1, \mbox{\bf R})
\Delta\nu (E_2, \mbox{\bf R + r}) \rangle \nonumber \\
& = & \lim_{\rho \to \infty}
\left\{ S_{\rho} (r, E_1, E_2) + S_{\rho} (r, -E_1, -E_2) \right.
\nonumber \\ & + & \left. R^+_{\rho} (r, E_1, E_2) + R^+_{\rho} (r,
-E_1, -E_2) \right\},
\end{eqnarray}
where $\Delta\nu$ is a fluctuating part of the density of states. It
is very easy to draw general conclusions about the singular part of
this expression. Indeed, in terms of the localized states we obtain
\end{multicols}
\widetext
\vspace*{-0.2truein} \noindent \hrulefill \hspace*{3.6truein}
\begin{eqnarray} \label{corr111}
& & K(r,E_1,E_2) = \left\langle \sum_{ij} \delta(\bbox{R} -
\bbox{R}_i) \delta(\bbox{R} + \bbox{r} - \bbox{R}_j)
\delta(\varepsilon_i - E_1) \delta (\varepsilon_j -
E_2)\right\rangle_c.
\end{eqnarray}
\hspace*{3.6truein}\noindent \hrulefill 
\begin{multicols}{2}
\noindent
If we now assume that there are no states with different energies
localized at the {\em same} space point ({\em i.e.} $\bbox{R}_i =
\bbox{R}_j$ implies $\varepsilon_i = \varepsilon_j$, this assumption
effectively serving as an ultraviolet cut-off), it is evident that
Eq. (\ref{corr111}) contains a singular part,
\begin{equation} \label{singpart1}
K_{sing} (r,E_1,E_2) = \nu(E_1) \delta(E_1 - E_2)
\frac{\delta(r)}{2\pi r},
\end{equation}
which originates from the term with $i = j$.
On the other hand, the remaining terms are {\em regular} in $r$, and
can be obtained perturbatively from the equations (\ref{basic2}) and
(\ref{basic3}), to which the initial conditions (\ref{indcond1}) must
be inserted in right-hand sides. Below we deal in this way with the
regular part $K_{reg}$ of the level-level correlation function
(\ref{corpert1}).

\subsection{Regular part, zero temperature}

For $T = 0$, the correlation function $S_{reg}(r, E_1, E_2)$ describes
fluctuations of the densities of states at the opposite sides of the
Fermi level, $E_1 E_2 < 0$; otherwise, it equals zero. The result for
this quantity is
\begin{equation} \label{cor11}
S_{reg}(r, E_1, E_2) =  - \nu_0^2
\theta(\rho^* - r).
\end{equation}
with the notation $\rho^* = \rho_{\vert E_1 \vert + \vert E_2\vert}$.
In the derivation of  (\ref{cor11}) we have used the identity
\begin{displaymath}
\int d\bbox{n} \frac{\delta( \vert \bbox{r} + \bbox{n}\rho
\vert)}{2\pi \vert \bbox{r} + \bbox{n}\rho \vert} = \frac{1}{r}
\delta(r - \rho).
\end{displaymath}
The function $R_{reg}^+ (r, E_1, E_2)$ is responsible for the correlations
of levels at the same side of the Fermi surface,
\begin{eqnarray} \label{cor12}
R_{reg}(r, E_1, E_2)  =   \frac{\nu_0^2}{2} \theta (\tilde \rho - r),
\end{eqnarray}
with $\tilde \rho = \min [\rho_{E_1}, \rho_{E_2}]$. Eq. (\ref{cor12})
applies for $E_1 > 0, E_2 > 0$ as well as $E_1 < 0, E_2 < 0$;
therefore we dropped the superscript $+$.

The first observation to be made is that the equations (\ref{cor11})
and (\ref{cor12}) taken together obey the ``detailed balance''
 relation,
\begin{eqnarray} \label{balance1}
& & \int_{-\infty}^{\infty} dE_1 dE_2 \ K_{reg} (r, E_1, E_2) \nonumber
\\
&= & 2 \int_{0}^{\infty} dE_1 \int_{-\infty}^0 dE_2 \ S_{reg}(r, E_1,
E_2) \nonumber \\
& + & 2 \int_{0}^{\infty} dE_1 \int_0^{\infty} dE_2
R_{reg}(r, E_1, E_2)  = 0,
\end{eqnarray}
{\em for any} $r$.

Obviously, $S_{reg}$ and $R_{reg}$ are zero without interactions.
Therefore, they express interaction-induced spatial correlations
between the sites. These correlations are  the stronger the closer the
two energies $E_1$ and $E_2$ are to the Fermi surface: though the
amplitudes of the correlators are constant, their radii, $\tilde\rho$
or $\rho^*$, increase as $E_1 \to 0$ {\em and} $E_2 \to 0$. The sign of
these correlations is negative (level repulsion) for the energies of
the opposite signs.  The origin of this repulsion is that the
transpositions, which build up the ground state of the interacting
system, involve pairs of sites across the Fermi surface. As a result,
in the ground state, when  the transpositions are
energetically unfavorable, the occupied and empty sites
tend to stay away from each other.

The positive sign of the correlator $R$ (level attraction) can be
understood as follows.  Close in space occupied sites ``feel''
the repulsion from  the {\em same} empty sites
and, hence, tend to gather closer together.
Note, that the phenomenon of ``clustering of  like sites'' was
identified in course of numerical simulations
with $V(\rho)\propto \rho^{-1}$ (and, thus, fully-developed Coulomb gap)
in  Ref. \onlinecite{Davis}.

\subsection{Regular part, finite temperatures}

The first order corrections to the functions $S$ and $R^+$ can be
actually calculated in a closed form for an arbitrary
temperature. Combining them into the level correlation function
according to Eq. (\ref{corpert1}), we obtain an expression
\end{multicols}
\widetext
\vspace*{-0.2truein} \noindent \hrulefill \hspace*{3.6truein}
\begin{eqnarray} \label{cortemp1}
K_{reg} (r, E_1, E_2) & = & \frac{\nu_0^2}{2} \Bigl\{ -2\psi[E_1 - E_2
+ V(r)] f_F (E_1) f_F (-E_2) - 2\psi[E_2 - E_1 + V(r)] f_F (-E_1) f_F
(E_2) \Bigr. \nonumber \\
& + & \left. \psi(E_1 - E_2) \left[ f_F (E_1) f_F (-E_2 - V(r)) + f_F
(-E_2) f_F (E_1 - V(r)) \right] \right. \nonumber \\
& + & \Bigl. \psi(E_2 - E_1) \left[ f_F (E_2) f_F
(-E_1 - V(r)) + f_F (-E_1) f_F (E_2 - V(r)) \right] \Bigr\}.
\end{eqnarray}
\hspace*{3.6truein}\noindent \hrulefill 
\begin{multicols}{2}
\noindent
This is a general result valid for an arbitrary relation between the
energies $E_1$, $E_2$, $V(r)$, and $T$. It is straightforward to check
that for $T = 0$ Eq. (\ref{cortemp1}) reproduces
expressions (\ref{cor11}) and (\ref{cor12}). Also, the detailed
balance relation,
\begin{displaymath}
\int dE_1 dE_2 K_{reg} (r, E_1, E_2) = 0,
\end{displaymath}
is fulfilled for any $r$ and $T$.

The expression (\ref {corpert1}) for the level-level correlator
can be presented in a compact form when the  energies $E_1$ and $E_2$
coincide,
\end{multicols}
\widetext
\vspace*{-0.2truein} \noindent \hrulefill \hspace*{3.6truein}
\begin{eqnarray} \label{compact}
K_{reg} (r,E,E) & = & \frac{\nu_0^2\sinh\Bigl(V(r)/2T\Bigr)}
{\cosh^2\left(E/2T\right)}
\Biggl[\frac{\cosh\Bigl(V(r)/2T\Bigr)\cosh^2(E/2T)+
\sinh\Bigl(V(r)/2T\Bigr)\sinh^2(E/2T)}{\cosh(E/T)+
\cosh\Bigl(V(r)/T\Bigr)}\Biggr]. 
\end{eqnarray}
\hspace*{3.6truein}\noindent \hrulefill 
\begin{multicols}{2}
\noindent
We see that the abrupt cut-off of the correlator (\ref{cor12}) at $T=0$
gets smeared at finite temperature. In the vicinity of the
Fermi level  ($\vert E \vert\ll T$)  Eq. (\ref {compact}) simplifies
to
\begin{equation} \label{part11}
K_{reg} (r,0,0) = \frac{\nu_0^2}{2} \tanh \left( \frac{V(r)}{2T}
\right).
\end{equation}
The above expression illustrates the decay of the correlator
with increasing either the distance $r$ or the temperature $T$.
In the opposite limit ($|E|\gg T$) the generalization of
(\ref{cor12}) to the finite temperatures takes the form
\begin{equation}
\label{bigE}
K_{reg} (r,E,E)=\nu_0^2\Biggl[\frac{\exp\Bigl(V(r)/T\Bigr)-1}
{\exp\Bigl(|E|/T\Bigr) + 2\cosh\Bigl(V(r)/T\Bigr)}\Biggr],
\end{equation}
which shows that at finite $r$, instead of the abrupt termination
of  (\ref{cor12}) at $|E|=V(r)$, the correlator decays with
departure of $E$ from the Fermi level as $\exp(-|E|/T)$.

Finally, let us  consider the behavior of the correlator
at  large distances ($V(r)\ll T$).
Expanding (\ref {compact}) in the first order in $V(r)/T$ yields
\begin{equation} \label{part12}
K_{reg} (r,E,E) = \frac{\nu_0^2 V(r)}{4T\cosh^2 (E/2T)},
\end{equation}
so that  [in accordance with Eq. (\ref {part11})]
$K_{reg} (r,E,E)$ decays  as $V(r)$.
The most important consequence of Eq. (\ref {compact}) is
that at any distances and temperatures the sign of the
level-level correlator at coinciding energies remains
positive (level attraction). Below we will see that this
is not the case for {\em arbitrarily} small mismatch $E_1-E_2$.
To illustrate this point we analyze several limiting cases for
which the general expression (\ref {cortemp1}) for
$K_{reg} (r, E_1, E_2)$  can be simplified.

{\bf Far tail in $r$, different energies}, $V(r) \ll \vert E_1 - E_2
\vert, T$. In this limit  we get
\begin{eqnarray} \label{part13}
K_{reg} (r,E_1,E_2) & = & -\frac{\nu_0^2 V^2(r)}{16T^2} \exp \left( -
\frac{\vert E_1 - E_2 \vert}{2T} \right) \nonumber \\
& \times & \left\{ \frac{1}{\cosh (E_1/2T) \cosh^3 (E_2/2T)}
\right. \nonumber \\
& + & \left. \frac{1}{\cosh (E_2/2T) \cosh^3 (E_1/2T)} \right\}.
\end{eqnarray}
We see that, in contrast to the case of coinciding energies,
the correlation function is
negative (level repulsion) and proportional to $V^2(r)$.

{\bf Far tail in energy}, $E_1 = 0$, $\vert E_2 \vert \gg T, V(r)$. In
the leading order in $\exp(-\vert E_2 \vert /T)$ we obtain
\begin{eqnarray} \label{part14}
& & K_{reg} (r,0,E_2) \nonumber \\ 
& = & -\frac{\nu_0^2}{2} \exp \left( \frac{V(r) - 2\vert
E_2 \vert}{2T} \right) \frac{\sinh^2 (V(r)/2T)}{\cosh (V(r)/2T)}.
\end{eqnarray}
This expression indicates that for large enough energy separation
the repulsion persists for arbitrary relation between $V(r)$ and $T$.

{\bf Crossover from short- to long-distance behavior}. Consider
$K_{reg} (r, E+\Delta/2, E-\Delta/2)$ with $T, \Delta \ll|E|$. We have
seen already that this function (at least for $\Delta \ll T$) is
positive for low distances $r$, see Eq. (\ref{part11}). We are now
going to show that, as a function of the separation $r$, this function
changes sign at $r = \rho_{\Delta}$ and becomes negative (level
repulsion) for long distances. Indeed, for $\rho_E \ll r <
\rho_{\Delta}$ ($\Delta < V(r) \ll |E|$) in the leading order we find
\begin{eqnarray} \label{part15}
& & K_{reg} (r, E+\Delta/2, E-\Delta/2) = \nu_0^2 \exp \left( \frac{\Delta
- 2|E|}{2T} \right) \nonumber \\
& \times & \left[ \exp \left( \frac{V(r) - \Delta}{T} \right)
- 1 \right],
\end{eqnarray}
which is positive and corresponds to the level attraction. The
level-level correlation function turns to zero for $r =
\rho_\Delta$. For longer distances, the leading order is proportional
to $\exp(-2|E|/T)$ and becomes ($r > \rho_{\Delta}$)
\begin{eqnarray} \label{part16}
& & K_{reg} (r, E+\Delta/2, E-\Delta/2) = -\nu_0^2 e^{-2|E|/T}
\nonumber \\
& \times & \left( 1 +
e^{-\Delta/T} \right) \left( e^{V(r)/T} - 1\right) \sinh \left(
\frac{V(r)}{T} \right),
\end{eqnarray}
which matches Eq. (\ref{part13}). The levels, indeed, repel each
other. We emphasize once more that the transition between level
attraction and level repulsion happens when $V(r) = \Delta$, which
corresponds to much longer distances than $V(r) = |E|$.

\subsection{Level number variance}

We now turn to the question about fluctuations of the number of levels
within a certain energy strip which was formulated in the Introduction.

For weakly disordered systems, the variance of the number of levels in
the energy strip of the width $E$ serves as quantitative measure
of the deviation of the level statistics from Poissonian  (see {\em
e.g.} Ref. \onlinecite{Mirlin}). Provided the strip is taken far from
the band edges and contains many levels, this {\em level number
variance} is a function of $E$ only, and  does not depend on the
absolute position of the strip.

Drawing an analogy with the weakly disordered case, we define the
quantity $\Sigma (\Omega, E_1, E_2)$, which is the variance of the
number of levels which are located within the spatial region
$\Omega$ and have energies between $E_1$ and $E_2$. This level number
variance can be trivially expressed through the level-level
correlation function,
\begin{eqnarray} \label{lnvdef1}
& & \Sigma (\Omega, E_1, E_2) \nonumber \\ 
& = & \int_{E_1}^{E_2} dE dE' \int_{\Omega}
d\bbox{R} d\bbox{R'} K(\vert \bbox{R} - \bbox{R'} \vert, E, E').
\end{eqnarray}
Unlike in weakly disordered case,  for many-electron system under
study the level number variance  depends on both  $E_1$ and $E_2$, not
merely on the difference $E_1 - E_2$.

Eq. (\ref{lnvdef1}) can be significantly simplified if we assume that
the spatial domain $\Omega$ is sufficiently large, so that for
relevant energies the correlator $K$ is very small outside this
domain. Then $\Sigma$ is essentially proportional to the volume of
$\Omega$. Defining the partial level number variance (referred below
as LNV), $\sigma(E_1, E_2) = \Omega^{-1} \Sigma (\Omega, E_1, E_2)$,
and using the decomposition of $K$ into regular and singular parts, we
obtain
\begin{equation} \label{lnvdef2}
\sigma(E_1, E_2)- \nu_0 (E_2 - E_1)  =    \sigma_{sing} (E_1, E_2) +
\sigma_{reg} (E_1, E_2),
\end{equation}
where $\sigma_{sing}$ and $\sigma_{reg}$ are defined as
\begin{equation} \label{lnvSing}
\sigma_{sing} (E_1, E_2)  =  \int_{E_1}^{E_2} dE \left[ \nu(E) -
\nu_0 \right],
\end{equation}

\begin{equation} \label{lnvReg}
\sigma_{reg} (E_1, E_2) =  \int_{E_1}^{E_2} dE dE' \int_0^{\infty}
2\pi r dr K_{reg} (r, E, E').
\end{equation}
The value $\nu_0 (E_2 - E_1)$ corresponds to non-interacting electrons;
$\sigma_{sing}$ and $\sigma_{reg}$ are the interaction-induced
corrections. The  physical meaning  of these corrections is
quite different. Correction $\sigma_{sing}$ accounts for
the net depletion of the strip due to reduction of the density
of states. It does not bear information about deviations from the
Poissonian statistics. These deviations are described  by
$\sigma_{reg}$.

Generally, both corrections $\sigma_{sing}$ and $\sigma_{reg}$ are
complicated functions  of $E_1$, $E_2$, $T$, and the form of the
potential. For this reason, below we consider only several instructive
particular cases.

{\bf $T = 0$, symmetric interval}. Suppose that the energy strip is
symmetric with respect to the Fermi level: $E_1=-E/2$, $E_2=E/2$. At
zero temperature the ``non-Poissonian'' contribution to LNV
can be evaluated by substituting Eqs. (\ref{cor11}) and (\ref{cor12})
into Eq. (\ref{lnvReg}),
\begin{eqnarray}\label{lnvMain}
& & \sigma_{reg} (-E/2, E/2) = \nonumber \\
& & 2\pi\nu_0^2\int_{0}^{E/2}dE_1
\int_{0}^{E/2}dE_2\int_0^{\infty}dr r \theta[\tilde\rho -r] \nonumber
\\ & - & 4\pi\nu_0^2 \int_{0}^{E/2}dE_1
\int_{-E/2}^{0}dE_2\int_0^{\infty}dr r \theta[\rho^{\ast} -r].
\end{eqnarray}
With the  use  of definitions of $\tilde\rho(E_1,E_2)$ and
$\rho^{\ast}(E_1,E_2)$, integration over $r$ and one of the
energies in Eq. (\ref{lnvMain}) can be performed explicitly.
Then  $\sigma_{reg} (-E/2, E/2)$ takes the concise form
\begin{equation} \label{lnv01}
\sigma_{reg} (-E/2, E/2) = -2\pi\nu_0^2 \int_{E/2}^E dE' (E - E')
\rho^2_{E'}.
\end{equation}
This expression is one of the central results of the present study.
Note that initial formula (\ref{lnvMain}) for $\sigma_{reg} (-E/2,
E/2)$ contains two competing terms. The first term is positive and
originates from attraction between the occupied as well as between
the empty levels. The second term is negative reflecting the
repulsion between the levels with different occupation numbers.
Eq. (\ref{lnv01}) shows that the second term wins, {\em  i.e.} the
{\em net} deviation from the Poissonian statistics for $T=0$ and
symmetric energy interval corresponds to the {\em repulsion}.
To proceed further, we choose for concreteness the potential
$V (\rho)=V_0\Bigl(a/\rho\Bigr)^{1/2}\exp(-\rho/a)$ ($\rho \gg a)$,
which pertains to the planar electron system between two gates.
At short distances, $\rho \ll a$, the potential is Coulomb
(see Section \ref{denssect}). Consider $E \ll V_0$. Then  the
integrals (\ref{lnvSing}) and (\ref{lnv01}) are determined by
distances $\rho_E \gg a$ and can be calculated explicitly. The result
for LNV reads
\begin{eqnarray} \label{lnv02}
& & \sigma (-E/2, E/2) - \nu_0 E \\
& = & -[\nu_0-\nu(0)] E + 2\pi \left( \ln 2 - 1/2 \right)
\nu_0^2 a^2 E^2 \ln \left( \frac{V_0}{E} \right),\nonumber 
\end{eqnarray}
where the difference $\nu_0-\nu(0)$ is equal to
$\pi^{3/2}\nu_0^2a^2V_0/2$ as follows from Eq. (\ref{dens3}).
It is seen from  Eq. (\ref{lnv02}) that the leading
effect of interactions on LNV comes from the depletion of the
strip. The effect of level-level correlations is smaller in
parameter $E/V_0 \ll 1$. Obviously enough, the
interaction-induced correction to LNV is much smaller than $\nu_0E$,
since $\nu_0V_0a^2 \ll 1$ is the expansion parameter of the
perturbation theory.

Note  that, in the  leading order, $\sigma_{reg} (-E/2, E/2)$
is proportional to $E^2 \ln^2 (V_0/E)$. However, this contribution
is canceled by the corresponding part in  $\sigma_{sing} (-E/2, E/2)$.
As a result, the second term in (\ref{lnv02}) contains $\ln (V_0/E)$
in the first power.

For $E \gg V_0$ the interaction-induced correction to LNV
is determined by the short-distance  behavior of
$V(\rho)$. It depends on ultraviolet cut-off required to
satisfy the sum rule $\int [\nu(E) - \nu_0] dE=0$
(see Section \ref{denssect}),  and is not discussed here.

{\bf High temperatures, symmetric interval}. The above conclusion
that correlations cause a negative correction to LNV does not
hold as the temperature is elevated. To illustrate this, we
choose the limit when the temperature is  much higher
than the width of the strip. In contrast to the  $T=0$ case considered
above, we will also assume that the strip is much wider than the
crossover energy $V_0 \sim e^2/\kappa a$. As it was already
mentioned, for $T \gg V_0$ the exact form of the  long-distance
fall-off of the potential appears to  be not essential.
More precisely, the characteristic
decay radius $a$ enters the results only under the logarithm, as we
established in Section \ref{denssect} while calculating the density of
states [see Eq. (\ref{L1})]. In this calculation,
high temperature introduced a
short-distance cut-off at $\rho \sim \rho_{T}=e^2/\kappa T$.
The major contribution to $[\nu(E) - \nu_0]$ came from the wide
interval of distances $\rho_{T}< \rho < a $ within which
the potential is Coulomb. It appears that the same  simplifications
are valid for  the calculation  of $\sigma_{reg}(-E/2,E/2)$.
Namely,  it is  safe to use the expansion with respect to
$V(r)/T$ in the  general formula  (\ref{cortemp1})
for the correlator $K_{reg}$. Substituting this expansion
into Eq.  (\ref{lnvReg}) and performing integration
first over $E_1, E_2$ and then over $r$ we obtain
\begin{equation}\label{lnvRegInter}
\sigma_{reg} (-E/2, E/2)=\frac{2\pi \nu_0^2 e^4 E}{\kappa^{2} T}
 \ln \Bigl(\frac{\kappa a E}{e^2}\Bigr),
\end{equation}
The positive sign of $\sigma_{reg}$ indicates that at high $T$
correlations cause a super-Poissonian behavior of LNV. The remaining
task is to compare Eq. (\ref{lnvRegInter}) with $\sigma_{sing}$.
The expression for $\sigma_{sing} (-E/2, E/2)$ immediately
follows from Eq. (\ref{ashigh1}).
\begin{equation}\label{lnvSingIntermed}
\sigma_{sing} (-E/2, E/2)= -\frac{(2\ln 2 -1)\pi e^4 \nu_0^2E}
{4\kappa^2 T} \ln \Bigl(\frac{\kappa a T}{e^2}\Bigr).
\end{equation}
Comparing (\ref{lnvRegInter}) and (\ref{lnvSingIntermed}),
we see that, besides different numerical coefficients, they
differ in logarithmic factors, which were assumed to be large
parameters in the course of the calculation. Since $T$ is much bigger
than $E$, the logarithmic factor in (\ref{lnvSingIntermed})
dominates. 

In both particular cases considered above,  contribution of
correlations to LNV was parametrically smaller than the
contribution from the change of the density of states. It is
interesting to find out whether or not the opposite relation is
possible within a certain range of parameters. Below we illustrate
that the opposite relation is indeed possible.

{\bf High temperatures, asymmetric interval}. Consider now the
case of strongly asymmetric (with respect to the Fermi level) energy
strip, $(E-\Delta/2, E + \Delta/2)$, with $E\gg\Delta$. The most
interesting situation for this case occurs when the conditions $V_0
\ll T \ll E$ are met. Then $[\nu(E)-\nu_0]$ is given by
Eq. (\ref{ashigh2}) and is small as $\exp(-E/T)$. Hence,
$\sigma_{sing}$ is also proportional to $\exp(-E/T)$. As a
result, unlike the symmetric-strip case, the interaction-induced
correction to LNV is dominated by $\sigma_{reg}$,
\begin{eqnarray}\label{lnva4}
& & \sigma(E-\Delta/2, E + \Delta/2)-\nu_0\Delta \nonumber \\
& = & \int_{E-\Delta/2}^{E+\Delta/2} dE dE' \int_0^{\infty}
2\pi r dr K_{reg} (r, E, E').
\end{eqnarray}
From Eqs. (\ref{part15}) and (\ref{part16}) we see that the level-level
correlation function for $V(r) \lesssim E$ is also exponentially
suppressed. Then the major contribution  to (\ref{lnva4}) comes from
the  short distances where $E \lesssim V(r)$. For such $r$ we
can replace $K_{reg} (r, E_1, E_2)$ by $K_{reg}(r,E)$ in
Eq. (\ref{lnva4}). Moreover, for $T\ll E \lesssim V(\rho)$
we  can use the zero-temperature result  (\ref{cor12}) for level-level
correlation function, namely $K_{reg} (r, E, E)= (\nu_0^2/2)
\theta(\rho_E-r)$. Finally, since the center of the strip, $E$, lies
well above $V_0$, the  parameter $\rho_E$ is determined by the Coulomb
part of the potential, {\em i.e} $\rho_E=e^2/\kappa E$. Combining all
together, we obtain for LNV
\begin{equation} \label{lnva2}
\sigma (E - \Delta/2, E + \Delta/2) = \nu_0 \Delta + \frac{\pi \nu_0^2
\Delta^2}{2E^2} \frac{e^4}{\kappa^2}.
\end{equation}
The super-Poissonian result for LNV is not surprising, since we
have already realized that like sites tend to attract each other.
To estimate the ratio of non-Poissonian and Poissonian contributions
to LNV, it is convenient to rewrite it as
$(\Delta/E)(\nu_0e^4/\kappa^2E)$. The first factor is small. Less
trivial is to realize that the second factor is also small. This
follows from the condition $E\gg T$, which we  exploited, and  from the
fact that in the perturbative regime the condition $T\gg
\nu_0e^4/\kappa^2$ should hold. As it was  mentioned above, this
condition ensures that the Coulomb gap is washed out by temperature.

\section{Conclusions} \label{concl}

In the present paper we have extended the self-consistent
description of the disordered interacting system of localized
electrons proposed in Ref. \onlinecite{mogil} to include the
correlation of the single-particle levels (both in energy and
in space). Correspondingly, our results allow to incorporate
correlations in the calculation of the density of low-lying neutral
excitations of the system (these excitations determine the specific
heat\cite{Baranovskii}). We obtained the analytical solution of the
self-consistent equations for the case when, due to the screening by
gates, interactions are weak enough to be treated perturbatively. Let
us discuss the conditions when the perturbative description is valid.

In the absence of screening, the width of the Coulomb gap can be
estimated by equating the low-energy asymptotics\cite{Efros75}
$\nu(E)\sim \kappa^2|E|/e^4$ of the density of states to the bare
density of states $\nu_0$. This yields $|E|\sim
\nu_0e^4/\kappa^2=E_C$. A perturbative solution applies when the
screening cuts off the Coulomb interaction at  short enough  distance,
$d$, so  that $V(d)\sim e^2/\kappa d \gg E_C$.
In this case the interaction-induced correction to the density
of states, $\delta \nu (E)$, is  much smaller than $\nu_0$ at {\em
all} energies. In other words, the perturbative regime corresponds to
a small enough $\nu_0$, which implies a broad distribution of the bare
energies of sites, {\em i.e.} to the limit of a strong external
disorder\cite{Xu}.

The above condition $d \ll \kappa/\nu_0e^2$ can be derived from a
very different consideration. Since the density of states at the Fermi
level is close to $\nu_0$ in the perturbative regime, the combination
$R_C=\kappa/\nu_0e^2$ formally represents the  static screening radius
in two dimensions. Then the condition of validity of the perturbative
description can be presented as $d \ll R_C$,  and has a
transparent interpretation that screening is not important for
short-range interactions.

At zero temperature and $E=0$ we obtained the perturbative result
$\delta\nu(0)/\nu_0 =  - \pi\nu_0 \int_0^{\infty}d\rho \rho V(\rho)$.
Note, that the r.h.s. of this expression can be interpreted as a
probability for an electron at the Fermi level to be transposed
as the interactions are switched on. Then the fact that
$\delta\nu(0)/\nu_0 \ll 1$ implies that the portion of sites
which are involved in transpositions ({\em i.e.} the sites, $i$, for
which the transposition with another site, $j$, is energetically
favorable) is relatively small. This ensures that the deviation from
the  Poissonian statistics of energy levels is weak.

Obviously, the perturbative description also applies when temperature
is high enough, $T\gg E_C$.

The predictions following from the theory developed, which are easiest
to test experimentally, concern the energy dependence of $\delta\nu$.
The precursor of the Coulomb gap should show up in the low-bias
behavior of the differential tunneling conductance $\delta G(U)\propto
\delta\nu(U)$. For tunneling between a metallic gate and
two-dimensional layer of localized electrons $\delta G(U)$ is
determined by Eq. (\ref {dens21}) (see also  Fig.~2) at $T=0$, and by
$F(U/T)$, where the function  $F$ is given by (\ref {func1}) and
plotted in Fig.~3, for high temperatures.

Let us discuss the physics left out within the self-consistent
approach. Firstly, while considering transpositions within pairs of
sites, we neglected the fact that,  upon each transposition
$i\rightarrow j$, the single-particle energies of all other sites
change. This change could trigger the ``secondary'' transpositions
which were forbidden (at $T=0$) before the transposition $i\rightarrow
j$ took place (polaronic effect \cite{selfconsis}). Secondly, we  were
concerned only with the question whether or not a transposition {\em
within a pair} is energetically favorable. The example given in
Ref. \onlinecite{polarons} shows that, even when all pair conditions,
$\Delta_{12}>0$ [ see Eq. (\ref {work1}) ] are satisfied, the work
required for {\em simultaneous} transfer of {\em two} electrons can be
negative (bipolaronic effect). Note, however, that both effects are
higher order in $\nu_0$ than the $\nu_0^2$-corrections captured by the
perturbative solution. For the same reason, we explicitly
disregarded triple correlations of the type $\langle \delta g \delta g
\delta g \rangle$ when deriving Eqs. (\ref{basic2}) and
(\ref{basic3}). Terms describing these correlations are also of the
order of $\nu_0^3$ and are beyond our precision.

We now give an overview of works discussing spectral statistics in
interacting systems. One avenue found in the literature is to study
(mostly numerically) the statistics of {\em many-particle}
levels. The early finding was that many-particle levels in
generic interacting systems obey Wigner-Dyson statistics, while
integrable ones are Poissonian distributed \cite{levels1}.
Subsequently, statistics of many-particle states have been discussed
in weakly disordered conductors \cite{levels2}, at the
metal-insulator transition \cite{levels3}, and in the insulating
regime \cite{Berkovits}, with the general conclusion that
the many-particle level distribution crosses from the non-interacting
one to the Poissonian one as the number of interacting particles
increases (thus, the statistics of three-particle states
are closer to the Poissonian form as those of two-particle states).
Another opportunity, realized in Refs. \onlinecite{PikusJETP} and
\onlinecite{PKS97}, is to study the addition spectrum of quantum dots
in the deeply insulating regime (defined as the increment of the
many-particle ground state energy when electrons are consecutively
added to the dot). 

Our results for level-level correlations correspond to yet another
experimental setup:  STM tip moves along the surface of the sample and
measures the tunneling conductance, proportional to the local density
of states\cite{gramada}. This is unrelated to the addition spectrum of
quantum dots, since to discuss the addition spectrum, one has to
account for the shift of the Fermi energy as electrons are added. What
is more, it is not clear whether the really measured addition spectrum
of such dots would be determined by the ground state energy rather
than by energies of the single-particle intermediate states, through
which an electron actually tunnels. What we discuss in this paper are
also not many-particle states but rather quasiparticle excitations
(corresponding to the poles of the single-particle Green's function);
this is an analog of the excitation spectrum peaks in the weakly
localized regime \cite{levels2} rather than many-body states which
compose each peak.

Finally, we remark that recently the issue of the Coulomb gap in {\em
quantum} interacting systems became a subject of several numerical
studies \cite{MacDonald,Eric,Epperlein,Choi,Shep1} in connection with
the quantum Hall effect and metal-insulator transition in two
dimensions. The results of our paper are strictly classical and are
valid in the deeply insulating regime, {\em i.e} we assumed that the
overlap integral between the localized wave functions is smaller than
the smallest energy scale in the problem; however, to match them to
the zero-bias anomaly in the weak disorder regime one inevitably has
to pass the quantum insulating domain \cite{Kopietz}.

\section*{Acknowledgments}

This work was supported  by the Swiss National Science Foundation
(Y.~M.~B.). M.~E.~R. acknowledges the support of the NSF under grant
INT-9815194. He is also grateful to the University of Geneva for
hospitality.  We thank The Aspen Center for Physics, where this work
was started, for hospitality and support.

\end{multicols}
\end{document}